\documentclass[aps,prd,reprint,showpacs,showkeys,preprintnumbers,superscriptaddress,nofootinbib]{revtex4-2}

\usepackage{graphicx,epsfig,color,xcolor}

\usepackage[english]{babel}
\usepackage{amssymb,amsfonts,amsmath,physics}
\usepackage{mathtools}
\usepackage{siunitx}
\usepackage{verbatim}
\usepackage{ulem}
\usepackage{soul}
\usepackage{float}
\usepackage{lipsum, babel}
\usepackage{acronym}
\usepackage{aas_macros}

\usepackage[colorlinks=true
,urlcolor=DARKBLUE
,anchorcolor=DARKBLUE
,citecolor=DARKBLUE
,filecolor=DARKBLUE
,linkcolor=DARKBLUE
,menucolor=DARKBLUE
,pagecolor=DARKBLUE
,linktocpage=true
,pdfproducer=medialab
,pdfa=true
,bookmarks=true
]{hyperref}

\newcommand{\be}{\begin{equation}} \newcommand{\ee}{\end{equation}}
\newcommand{\bea}{\begin{eqnarray}} \newcommand{\eea}{\end{eqnarray}}

\definecolor{MONZA}{HTML}{CF000F}
\definecolor{DARKBLUE}{HTML}{00008b}
\definecolor{DARKMAGENTA}{HTML}{8b008b}
\definecolor{DARKCYAN}{HTML}{008B8B}
\definecolor{DARKORANGE}{HTML}{FF8C00}
\definecolor{OBSERVATORY}{HTML}{049372}
\definecolor{GREENBAMBOO}{HTML}{006442}
\definecolor{TURQUOISE}{HTML}{36D7B7}
\definecolor{JUNGLEGREEN}{HTML}{26C281}

\begin{document}

\title{Threshold for PBH formation in the type-II region and its analytical estimation}

\author{Albert Escrivà}
\email{escriva.manas.alberto.k0@f.mail.nagoya-u.ac.jp}
\affiliation{Institute for Advanced Research, Nagoya University, \\
Furo-cho Chikusa-ku, Nagoya 464-8601, Japan}
\affiliation{Department of Physics, Nagoya University, \\
Furo-cho Chikusa-ku, Nagoya 464-8602, Japan}


\begin{abstract}
We numerically simulate the formation of Primordial Black Holes (PBHs) in a radiation-dominated Universe under the assumption of spherical symmetry, driven by the collapse of adiabatic fluctuations, for different curvature profiles $\zeta$. Our results show that the threshold for PBH formation, defined as the peak value of the critical compaction function $\mathcal{C}_{c}(r_m)$ (where $r_m$ is the scale at which the peak occurs), does not necessarily asymptotically saturate to its maximum possible value in the type-I region for sufficiently sharp profiles. Instead, the threshold is found in the type-II region with $\mathcal{C}_{c}(r_m)$ being a minimum. We find, for the cases tested, that this is a general trend associated with profiles that exhibit extremely large curvatures in the linear component of the compaction function $\mathcal{C}_{l}(r) \equiv -4r \zeta'(r)/3$ shape around its peak $r_m$ (spiky shapes). To measure this curvature at $r_m$, we define a dimensionless parameter, $\kappa \equiv -r^{2}_m \mathcal{C}_l''(r_m)$, and we find that the thresholds observed in the type-II region occur for sufficiently large $\kappa$ for the profiles we have used, contrary to expectations. By defining the threshold in terms of $\mathcal{C}_{l,c}(r_m)$, we extend previous analytical estimations to the type-II region, which is shown to be accurate within a few percent when compared to the numerical simulations for the tested profiles. Our results suggest that current PBH abundance calculations for models where the threshold lies in the type-II region may have been overestimated due to the general assumption that it should saturate at the boundary between the type-I and type-II regions.
\end{abstract}
\pacs{
}
\maketitle
\acresetall

\acrodef{GW}{gravitational wave}
\acrodef{PT}{phase transition}
\acrodef{SC}{smooth crossover}
\acrodef{SM}{Standard Model}
\acrodef{QCD}{Quantum Chromodynamics}
\acrodef{EW}{electroweak}
\acrodef{CMB}{cosmic microwave background}
\acrodef{PBH}{primordial black hole}
\acrodef{DM}{Dark Matter}
\acrodef{FLRW}{Friedmann‐-Lema\^itre--Robertson--Walker}
\acrodef{MS}{Misner--Sharp}

\section{Introduction}
Primordial Black Holes (PBHs) \cite{Zeldovich:1967lct, 10.1093/mnras/152.1.75, Carr:1974nx} are black holes that may have formed in the early Universe without a stellar origin, through various mechanisms (see for reviews \cite{Khlopov:2008qy,Sasaki:2018dmp,Carr:2020gox,Green:2020jor,Carr:2020xqk,Escriva:2022duf}). The most widely studied scenario involves the collapse of super-horizon curvature fluctuations (the subject of this study), particularly during a radiation-dominated era \cite{Carr:1975qj}. PBHs remain a promising candidate to explain a significant fraction of dark matter, especially in the asteroidal mass range, with $M_{\rm PBH} \in [10^{-15}, 10^{-10}] M_{\odot}$ \cite{Chapline:1975ojl}.

A common approach to studying the PBH formation process is the assumption of spherical symmetry (see \cite{Escriva:2021aeh} for a review focusing on numerical results), which is based on the fact that, with Gaussian statistics, large peaks \cite{1986ApJ...304...15B} (those that may significantly contribute to the production of a large quantity of PBHs) are approximately spherical. As a result, gravitational collapse is assumed to be spherical. This assumption has recently been tested for the case of Gaussian statistics with a monochromatic power spectrum \cite{Escriva:2024aeo,companion}. For the remainder of this work, we will thus assume spherical symmetry, although in specific scenarios, non-sphericities may play a role.

In estimating the abundance of PBHs in our Universe, it has been crucial to develop precise statistical methods to account for PBH production, with peak theory \cite{1986ApJ...304...15B} being a common approach \cite{Germani:2018jgr,Yoo:2018kvb,Germani:2019zez,Young:2019yug,Yoo:2019pma,Yoo:2020dkz,Young:2020xmk,Yoo:2022mzl,Pi:2024ert,Fumagalli:2024kxe}. One of the most important quantities in determining the abundance of PBHs is the threshold of black hole formation , which in our settings is defined at super-horizon scales\footnote{Specifically, we will characterize the threshold for PBH formation using a variable constructed from the curvature fluctuation $\zeta$, which remains time-independent on super-horizon scales.} (when adiabatic fluctuations are frozen and statistical methodologies can be used), as PBH production is highly sensitive (in particular with exponential dependence \cite{Carr:1975qj}) to the conditions that lead to the formation of these black holes. Typically, relativistic numerical simulations are necessary to study the highly nonlinear behavior of the gravitational collapse and infer which specific curvature profiles at superhorizon scales will form black holes or not. The quantity that characterizes the threshold for PBH formation has been the subject of much debate \cite{Shibata:1999zs,Niemeyer:1997mt,Musco:2004ak,Nakama:2013ica,Harada:2013epa}. However, as pioneeringly mentioned in Ref.\cite{Shibata:1999zs}, the peak value (the maximum) of the compaction function $\mathcal{C}(r_m)$ (which will be defined later in the comoving gauge) was identified as a quantity that may be useful for characterizing the threshold. This observation has been considered in several later numerical works in different scenarios \cite{Harada:2015yda,Musco:2018rwt,Escriva:2019phb,Musco:2020jjb,Escriva:2020tak}. Reference \cite{Escriva:2019phb} found, for several fluctuation shapes characterized by isolated peaks of $\mathcal{C}(r)$, that the profile dependence of $\mathcal{C}_c(r_m)$ is mainly sensitive to its curvature at the maximum $r_m$. This, together with the fact that the averaged critical compaction function was found to have an approximately universal value $\bar{\mathcal{C}}_{c} \approx 2/5$ for a set of standard curvature profiles, allowed for the creation of an analytical formula to predict $\mathcal{C}_c(r_m)$ accurately taking into account the profile dependence, with only a few percentages of deviation compared to the numerical results of \cite{Escriva:2019nsa}. This is essential for accounting for the profile dependence of curvature fluctuations and differentiating the significance of different models for PBH abundance estimates, without relying on extensive numerical simulations, and avoiding unrealistic estimates that arise from assuming the same threshold of black hole formation for all profiles.

Numerical studies \cite{Musco:2018rwt,Escriva:2019nsa} have shown that $\mathcal{C}_{c}(r_m) \in [2/5, 2/3]$ in a radiation-dominated era for critical initial data corresponding to type-I fluctuations, i.e., fluctuations where the areal radius is a monotonic increasing function, and where $\mathcal{C}_{c}(r_m)$ corresponds to the maximum value of the mass excess over the areal radius at super-horizon scales. On the other hand, recent studies (see \cite{Uehara:2024yyp} for the new type A/B PBH classification) have shown that for specific curvature profiles in models with large negative non-Gaussianities \cite{Shimada:2024eec,Inui:2024fgk}, as well as in exponential-shaped profiles \cite{MIO}, the threshold for PBH formation is found in the type-II region (the critical $\zeta_{c}(r)$ corresponds to type-II fluctuations). This corresponds to fluctuations where the areal radius is a non-monotonic function \cite{PhysRevD.83.124025}. In these cases, the threshold, defined as the peak of the compaction function, does not saturate at the maximum value of $2/3$, but instead decreases, and $\mathcal{C}_{c}(r_m)$ becomes a local minimum. This is contrary to expectations in the literature, which suggest that type-II fluctuations always collapse, forming black holes. This may raise the question of whether this is a generic and common phenomenon, which scale and parameterization can be used to characterize the critical initial conditions and how the thresholds, expressed in terms of compaction function quantities, behave for type-II critical data.

Existing analytical estimates \cite{Escriva:2019phb} (see also \cite{Musco:2020jjb}, which is effectively equivalent and based on \cite{Escriva:2019phb}) lie outside the regime of validity for type-II critical data. This is because it assumes that type-II fluctuations always collapse, with the threshold determined by the local maximum of the compaction function $\mathcal{C}(r)$ at $r_m$ in the type-I region (where the critical $\zeta_{c}(r)$ corresponds to type-I fluctuations). This highlights the need to extend the analytical framework to the type-II region and find a suitable parametrization to characterize the threshold in the type‑II region.

In this paper, we address these subjects using results from relativistic numerical simulations with different curvature profiles, with the aim of improving the understanding of the critical threshold conditions for type-II fluctuations in a parameter range not previously explored. Throughout the paper, we use geometrized units with $G = c = 1$.

\section{Cosmological set up for PBH formation}
We consider a Universe described by a \ac{FLRW} background filled with a perfect fluid characterized by an equation of state \( p = w\rho \), where \( p \) is the pressure, \( \rho \) is the energy density, and \( w \) is the equation-of-state parameter, that we fix to $w=1/3$. The energy-momentum tensor of the fluid is given by
\begin{equation}\label{eq:Tmunu_tensor}
    T_{\mu\nu} = (\rho + p) u_\mu u_\nu + p\, g_{\mu\nu},
\end{equation}
where \( u_\mu \) denotes the four-velocity of the fluid.

To incorporate spherical symmetry, we consider the following metric $g_{\mu \nu}$:
\begin{equation}\label{eq:metric455}
    \dd s^2 = -A(r,t)^2\,\dd t^2 + B(r,t)^2\,\dd r^2 + R(r,t)^2\, \dd\Omega^2\,,
\end{equation}
where \( \dd\Omega^2 \) is the metric on a unit two-sphere. In this coordinate system, \( t \) corresponds to the cosmic time, \( A(r,t) \) is the lapse function, and \( R(r,t) \) denotes the areal radius. With the above ansatz, the equations that describes the graviational collapse in the comoving gauge are given by the Misner-Sharp equations \cite{PhysRev.136.B571}. Using this metric, one can define the Misner–Sharp mass as the mass enclosed within a spherical surface of constant areal radius \( R(r,t) \):
\begin{equation}
    M(R) = \int_0^R 4\pi \rho\, \tilde{R}^2 \, \dd\tilde{R}\,.
\end{equation}

A key quantity in the study of PBH formation is the compaction function \cite{Shibata:1999zs} (see also \cite{Harada:2023ffo,Harada:2024trx} for a recent discussion), this corresponds to twice the mass excess over the areal radius. It is defined in the comoving gauge as
\begin{equation}
    \mathcal{C}(r,t) = 2\frac{M(r,t) - M_b(t)}{R(r,t)}\,,
\end{equation}
where \( M_b = 4 \pi \rho_b R^3/3 \) is the mass of the corresponding homogeneous FLRW background and $\rho_b$ its energy-density. 

\section{Curvature profiles and new parameterization}

\begin{figure}
\centering
\includegraphics[width=0.45\textwidth]{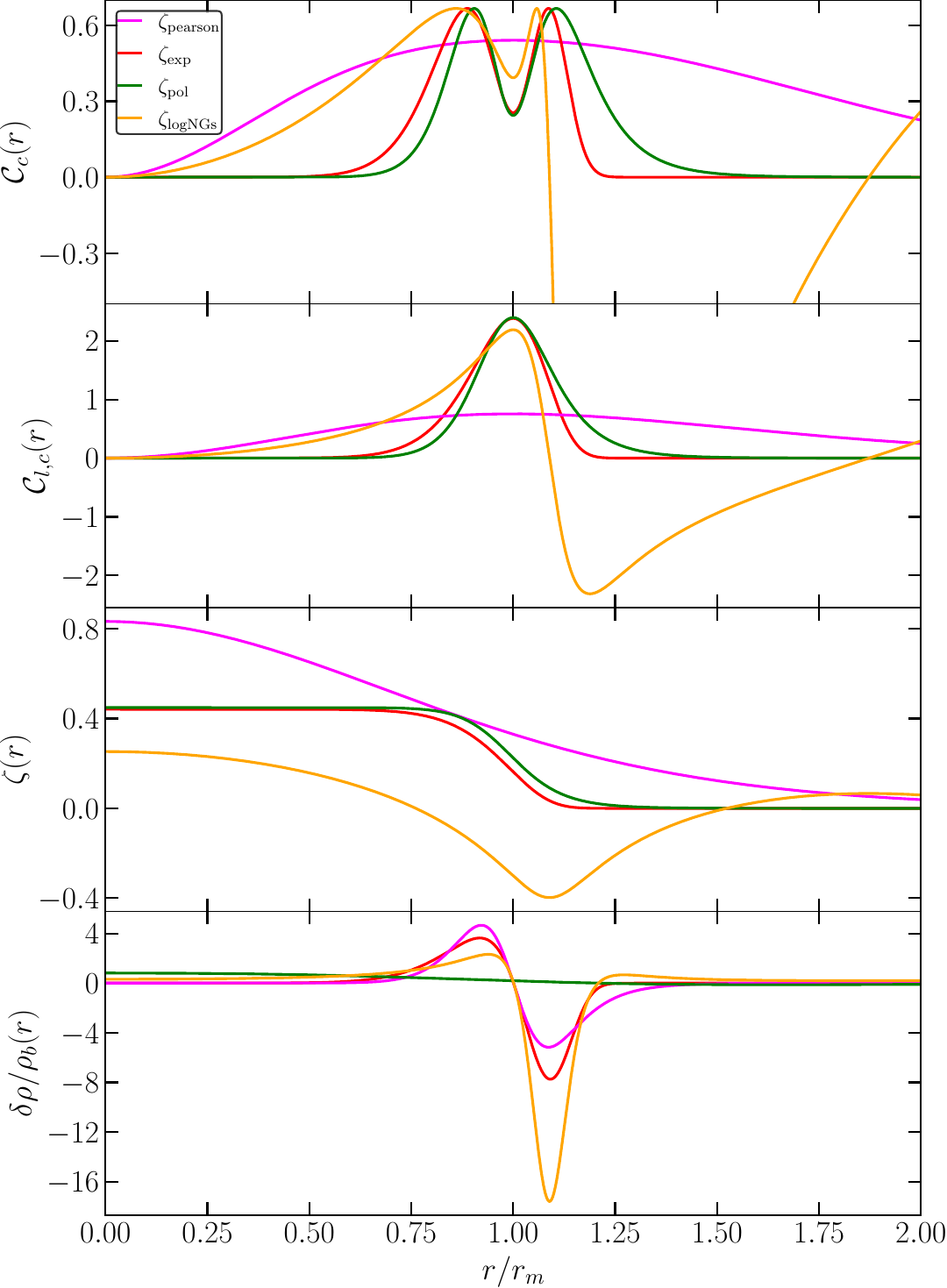}
\caption{Some profiles for the critical compaction function $\mathcal{C}(r)$ (first-panel), the linear component $\mathcal{C}_l(r)$ (second-panel), the $\zeta(r)$ profiles (third-panel) and the density contrast $\delta \rho/\rho_b(r)$ extrapolated at horizon crossing (fourth-panel)}.
\label{fig:shapes_C_Cl}
\end{figure}

In our study, we assume that PBH formation arises from the collapse of super‑horizon curvature fluctuations produced during the inflationary epoch. At early times, perturbations have a physical wavelength $L$ much larger than the Hubble radius $H^{-1} = a/\dot{a}$ with $a$ the scale factor of the FLRW background. Hence, we consider the long-wavelength approximation \cite{Lyth:2004gb} to determine the form of our initial metric and hydrodynamical variables (see \cite{MIO} for the exact expressions and details). This is based in
expanding the exact solutions in a power series of a parametter $\epsilon \equiv 1/(H(t)L(t))$ to the lowest non-vanishing order in $\epsilon^2(t) \ll 1$ to relate both scales \cite{Polnarev:2006aa,Harada:2015yda}. The spacetime metric at zero order in the gradient expansion with the presence of a super-horizon curvature fluctuation $\zeta$ and under the assumption of spherical symmetry is given by \cite{Shibata:1999zs}
\begin{equation} 
ds^2 = -dt^2 + a^2(t) e^{2\zeta(r)} \left( dr^2 + r^2 d\Omega^2 \right),
\label{metric2} 
\end{equation}
with $d\Omega^2 = d \theta^2+ \sin^2(\theta) d\phi^2$ and $r$ the conformally flat radial coordinate. The length-scale $L$ is then defined as $L=R_m=a r e^{\zeta}$. Curvature fluctuations $\zeta$ admit two different classifications: type-I corresponds to fluctuations with a monotonic increasing function for the areal radius $R= a r e^{\zeta}$. Whereas type-II corresponds to a non-monotonic $R$, satisfying that there exists a region where $R'<0$ \cite{PhysRevD.83.124025}. A compaction function $\mathcal{C}(r)$ can be defined \cite{Shibata:1999zs} (see \cite{Harada:2023ffo,Harada:2024trx} for a recent discussion). 
The compaction function at leading order in gradient expansion reads as \cite{Harada:2015yda}
\begin{equation}
\mathcal{C}(r) = \frac{2}{3}\left[1-(1+r \zeta')^2\right]= \mathcal{C}_l(r)-\frac{3}{8}\mathcal{C}^2_l(r),
\label{eq:compaction_function}
\end{equation}
where $\mathcal{C}_{l}(r)\equiv -(4/3) r \zeta'(r)$ is the linear component of the compaction function\footnote{This is also tipically defined as the linear density perturbation}. Notice that $\mathcal{C}(r)$ is a time-independent quantity since $\zeta(r)$ is frozen at super-horizon scales. This is essential for inferring the conditions for black hole formation at super-horizon scales and making the corresponding statistics of PBH production. Then we can write the linear component of the compaction function $\mathcal{C}_l(r)$ in terms of $\mathcal{C}(r)$ as 
\begin{equation}
    \mathcal{C}^{\pm}_l(r) = \frac{4}{3}\left( 1 \pm \sqrt{1-\frac{3}{2} \mathcal{C}(r)} \right)
    \label{eq:C_l},
\end{equation}
where the solution with $-,+$ corresponds to type-I/II fluctuations, respectively. In a radiation-dominated Universe, $\mathcal{C}_{c} \equiv \mathcal{C}_{c}(r_m)$ in the type-I region runs from $\mathcal{C}_c \in [2/5,2/3]$ \cite{Musco:2018rwt,Escriva:2019phb}, whereas in the type-II region it starts from $2/3$, decreases, and seems unbounded, as shown and discussed in \cite{MIO}. We can translate this range into the variable $\mathcal{C}_{l,c} \equiv \mathcal{C}_{l,c}(r_m)$, which gives $\mathcal{C}_{l,c}\in [(4/3)(1-\sqrt{2/5}) \, , \, 4/3]$ for the type-I, and for type-II starts from $\mathcal{C}_{l,c}=4/3$ and increases, for what currently we can not specify the existence of an upper bound. Then, $\mathcal{C}_{l,c}$ is always a monotonic increasing function in terms of $\mathcal{C}_{c}$ for the type-I/II regions, and therefore, it is convenient to define the threshold of PBH formation in terms of $\mathcal{C}_{l,c}$ to cover the type-II region. Indeed $\mathcal{C}_{l,c}$ is always a local maximum since $\zeta'(r_m)+r\zeta''(r_m)=0$ for both type-I/II regions and $-\mathcal{C}_{l,c}''= \pm \mathcal{C}''_{c}/\sqrt{1-3\mathcal{C}_{c}/2}$ (see Fig.\ref{fig:shapes_C_Cl}).

On the other side, the analytical formula done in \cite{Escriva:2019phb} allows us to correctly predict $\mathcal{C}_c$ in the type-I region (its regime of validity, focusing on local maxima $\mathcal{C}(r_m)$), this is done with the dimensionless parameter $q$ (introduced in \cite{Escriva:2019phb}), which measures the curvature of the shape around the peak of the compaction function
\begin{equation}
q  = \frac{-\tilde{r}^2_m \mathcal{C''}(\tilde{r}_m)}{4\mathcal{C}(\tilde{r}_m)} = \frac{-r^2_m \mathcal{C}''(r_m)}{4 \mathcal{C}(r_m)(1-3\mathcal{C}(r_m)/2)}
\label{eq:q_equation},
\end{equation}
and the analytical threshold $\delta_c(q) \equiv \mathcal{C}_{c}(r_m)$ is given by
\begin{equation}
    \delta_{c}(q)=\frac{4}{15}e^{-\frac{1}{q}}\frac{q^{1-\frac{5}{2q}}}{\Gamma\left(\frac{5}{2q}\right)-\Gamma\left(\frac{5}{2q},\frac{1}{q}\right)}
    \label{eq:delta_c},
\end{equation}
where $\Gamma(x)$ is the gamma function and $\Gamma(x,y)$ the incomplete gamma function. The $\tilde{r}$ corresponds to the areal radial coordinate, which makes the spacetime metric of Eq.\eqref{metric2} resemble the flat FLRW metric with a non-homogeneous curvature $K(\tilde{r})$ (see \cite{Harada:2015yda} for the transformation between $K(\tilde{r}), \zeta(r)$),
\begin{equation}
    d\Sigma^2 = a^2 \left(\frac{d\tilde{r}^2}{1-K(\tilde{r})\tilde{r}^2}+\tilde{r}^2 d\tilde{\Omega}^2\right).
    \label{eq:mtric_K}
\end{equation}
In this metric (for which type-II fluctuations cannot be realized \cite{PhysRevD.83.124025}), the compaction function reads as $\mathcal{C}(\tilde{r})=\frac{2}{3}K(r)r^2$, with maximum value given by $K(\tilde{r}_m)\tilde{r}^{2}_m=1 \Rightarrow \mathcal{C}(\tilde{r}_m)=2/3$, corresponding to the coordinate singularity\footnote{We note that the Separate Universe issue \cite{10.1093/mnras/168.2.399,PhysRevD.71.104009} was resolved in \cite{PhysRevD.83.124025}, where it was demonstrated that a proper coordinate transformation removes the apparent difficulty in the areal radial coordinate. Moreover, their analysis showed that the Separate Universe condition corresponds to the limit in which the curvature perturbation $\zeta$ diverges, which in our case does not constitute a problem.} in Eq.\eqref{eq:mtric_K}. On the other hand, the curvature $\mathcal{C}''(r_m)$ transitions from negative (type-I region) to positive (type-II region) values and the parameter $q$ diverges when $\mathcal{C}(r_m)=2/3$ (the marginal case when $\mathcal{C}''(r_m)=0$) due to the factor in the denominator.

To overcome this, let us consider another parameter. First, we know from \cite{Escriva:2019phb} that for critical initial data corresponding to type-I fluctuations the relevant profile-quantity that accurately determines $\mathcal{C}_c$ is the curvature shape of $\mathcal{C}$ at the scale $r_m$, and in particular its second derivative $\mathcal{C}''(r_m)$ (see Eq.\eqref{eq:q_equation}). In terms of the linear compaction function we have $-r^2_m\mathcal{C}''(r_m) = -r^2_m \mathcal{C}_l''(r_m)\left(1-3 \mathcal{C}_l(r_m)/4\right)$, which shows that both second derivatives are related. Second, in Fig.\ref{fig:shapes_C_Cl} (see also \cite{MIO}), we observe that the critical shapes $\mathcal{C}_{c}(r)$ for type-II fluctuations are characterized by spiky valley shapes at $r_m$, and for a sufficiently large gradient of the density contrast at leading order in the gradient expansion, the threshold of formation is found in the type-II region. Extrapolating the leading order in gradient expansion to the horizon crossing (see \cite{Musco:2018rwt} for details), the dimensionless derivative of $\delta \rho/\rho_b \equiv (\rho-\rho_b)/\rho_b$ evaluated at $r_m$ can be written as
\begin{equation}
-r_m\frac{\delta \rho'(r_m)}{\rho_b} = \frac{12}{18} \Bigl[\frac{\kappa}{2}+\mathcal{C}_{l}(r_m)-\frac{9}{8}\mathcal{C}^{2}_{l}(r_m)+\frac{9}{32}\mathcal{C}^{3}_{l}(r_m)\Bigr],
\end{equation}
where we have defined $\kappa \equiv -r^2_m \mathcal{C}''_l(r_m)$. For sharp shapes in $\delta \rho/ \rho_b$ at $r_m$, the term $\kappa$ dominates over the others, and therefore $\kappa$ may be the main parameter that needs to be considered. These two observations motivate the use of $\kappa$ as an alternative parameter to $q$ for characterizing different profiles in the type-II region, which accounts for the curvature of $\mathcal{C}_{l}(r)$ at its peak. Notice that the parameters $q$ and $\kappa$ can be related each other by (see for instance \cite{Germani:2023ojx}, where the variable $\kappa$ is used to include Eq.\eqref{eq:delta_c} in the non-linear statistics with type-I fluctuations)
\begin{equation}
\kappa = 4 \, q \,  \delta_c(q)\sqrt{1-3 \delta_c(q)/2}.
\label{eq:s_q}
\end{equation}
Let us now consider a set of different curvature profiles. In this work, we aim to analyze both standard and realistic curvature profiles characterized by an isolated dominant peak in the linear compaction function. Our motivation is to consider shapes with a functional form similar to those analyzed in previous studies (in particular \cite{Musco:2018rwt,Escriva:2019phb,Musco:2020jjb}) which focused on type-I fluctuations and extend the study to type-II fluctuations. For this purpose, we consider an exponential $\zeta_{\rm exp}$, polynomial $\zeta_{\rm pol}$ and pearson distribution $\zeta_{\rm pearson}$ shapes. The threshold behavior in terms of $q$ for these profile templates has also been shown to accurately describe the thresholds associated with typical profile shapes \cite{1986ApJ...304...15B}, which result from various power spectra and statistical methodologies; see, for instance, \cite{Musco:2020jjb,Yoo:2020dkz,Pi:2024ert}. We also consider a logarithmic non-Gaussian $\zeta_{\rm log,NG}$, which corresponds to the typical shape profiles (statistically motivated by peak theory \cite{1986ApJ...304...15B}) sourced by a monochromatic power spectrum in the presence of logarithmic non-Gaussianities \cite{Pi:2022ysn},
\begin{align}
\label{eq:profile_zeta_exp}
\zeta_{\rm exp}(r) &= \mu \exp(-(r/r_m)^{2\beta}),\\
\label{eq:profile_zeta_pol}
\zeta_{\rm pol}(r) &=  \frac{\mu}{1+(r/r_m)^{2 p}},\\
\label{eq:profile_zeta_pearson}
\zeta_{\rm pearson}(r) &=  \frac{\mu}{\left( 1+(r/(\sqrt{\alpha}\,r_m))^2\right)^\alpha}, \\
\label{eq:profile_zeta_log}
\zeta_{\rm log NGs}(r) &= -\frac{1}{\lambda}\log\left( 1-\lambda \,\zeta_G\right), \,\\
\nonumber
\zeta_G &= \mu \, \textrm{Sinc}\left(\frac{k_{*} r}{r_m(\mu,\lambda)}\right),
\end{align}
where $\mu$ is the peak value of $\zeta$ ($\zeta_G$ for the case $\zeta_{\rm log NGs}$), $k_{*}$ is the wave mode where the peak of a monochromatic power spectrum is located and $\beta,p,\alpha,\lambda$ the different parameters that modulate the shape. Notice that for the non-Gaussian profile case $r_m(\mu,\lambda)$ is dependent on $\mu,\lambda$ due to the non-Gaussianity $\lambda$. See Fig.\ref{fig:shapes_C_Cl}, where the critical profiles for some parameter cases are plotted. We can rewrite the profiles in terms of the new parametrization $\mathcal{C}_{l}(r_m)$ and $\kappa$, which can be found in Appendix~\ref{ref:apendix_shapes_parametrization}.

\section{Numerical results}
We use SPriBHoS-II code \cite{MIO,codigo_albert} to numerically compute the threshold for PBH formation in both the type-I and type-II regions for the profiles given by Eqs.\eqref{eq:profile_zeta_exp}-\eqref{eq:profile_zeta_pearson}. The code is based on a new methodology using the Misner-Sharp formalism, which can handle type-II simulations and avoid issues related to type-II fluctuations. This allows us to explore parameter ranges and profiles that were previously not considered.

We also take the numerical threshold values for Eq.\eqref{eq:profile_zeta_log} for $\lambda < 0 , \, \lambda \geq 0$ computed in \cite{Atal:2019erb, Shimada:2024eec}.

Specifically, once the curvature profile shape $\zeta$ is fixed, we use the long-wavelength solution at first order in the gradient expansion to set the hydrodynamic initial conditions (see \cite{MIO}). For cases with large curvatures $\kappa$, we increase the initial length scale $R_m$ relative to the Hubble scale in order to remain within the validity of the long-wavelength approximation. To ensure this, we use the Hamiltonian constraint equation, defined as $\mathcal{H} \equiv (M' - 4\rho R^2 R')/B$, to monitor the convergence of our simulations. We also verify that $\epsilon \lesssim 0.02$ is sufficient to guarantee convergence of the quasi-homogeneous solution for the cases tested\footnote{In particular, for the largest values of $\kappa$ considered, we find no change in the threshold values (computed with an abolute resolution of $\mathcal{O}(10^{-3})$) when reducing to smaller values $\epsilon$ than $\epsilon \lesssim 0.02$.}, we refer the reader to Appendix~\ref{ref:section_convergence} for details. We then apply a bisection method to track the formation of trapped surfaces (i.e., black hole formation) and determine the critical amplitude $\mu_c$. This value is subsequently converted into the corresponding compaction threshold $\mathcal{C}_{l,c}$. We refer the reader to \cite{MIO} for further details on the numerical implementation and methodology.

\begin{figure} 
\centering
\includegraphics[width=0.4\textwidth]{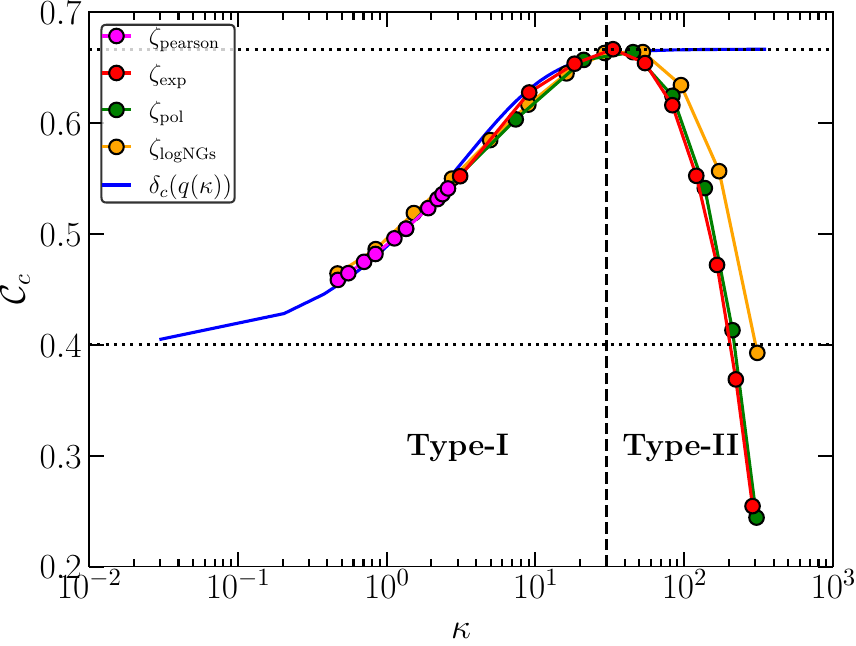}
\includegraphics[width=0.4\textwidth]{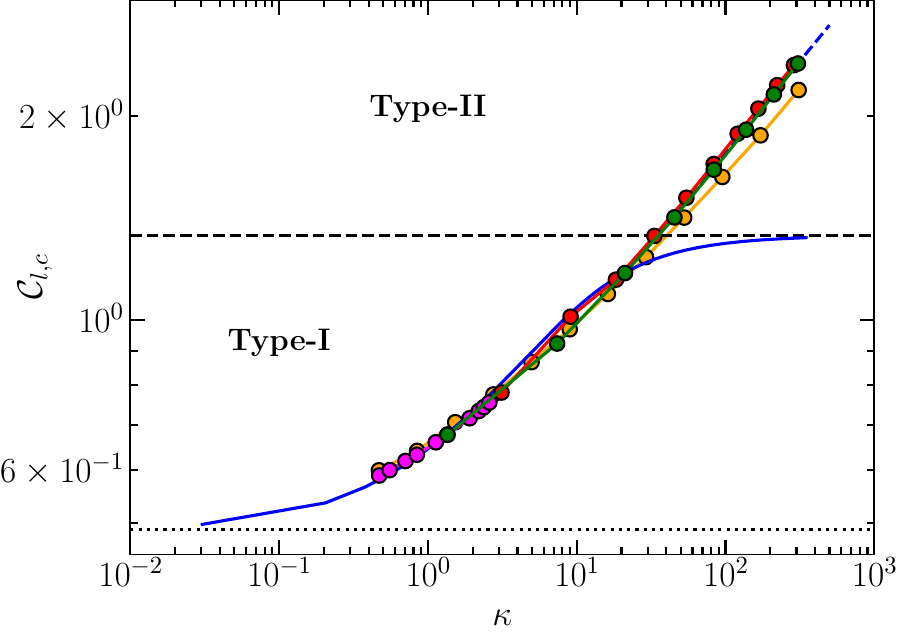}
\caption{Top-panel: Threshold in terms of $\mathcal{C}_{c}$, the bottom and top horizontal dotted lines correspond to $2/5$ and $2/3$, respectively, and the vertical dashed line is located at $\kappa = 30$. Bottom-panel: Threshold in terms of the linear component $\mathcal{C}_{l,c}$, the bottom and top horizontal dotted lines correspond to $(4/3)(1-\sqrt{2/5})$ and $4/3$ respectively, whereas the horizontal dashed line coincides with the top dotted one. The blue solid line in both panels corresponds to the analytical estimate of Eq.\eqref{eq:delta_c}. The dashed blue line corresponds to the numerical fitting of Eq.\eqref{eq:extension_analytical_estimate}.}
\label{fig:numresults_C_Cl}
\end{figure}

In Fig.\ref{fig:numresults_C_Cl}, we present the numerical results for $\mathcal{C}_{c}$ and $\mathcal{C}_{l,c}$ as functions of the parameter $\kappa$. The bottom and top dotted horizontal lines represent the minimum and maximum thresholds for the type-I region in both cases respectively, and the dashed line marks the boundary separating the type-I and type-II regions.
A key observation is that for all profiles considered, except the one in Eq.\eqref{eq:profile_zeta_pearson}\footnote{This is because $\alpha$ should be $\alpha>0$, and taking into acocunt that $\kappa/\mathcal{C}_{l}(r_m)=4\alpha/(1+\alpha)$ (see appendix Eq.\eqref{eq:appendix_pearson}), the maximum $(\kappa/\mathcal{C}_{l})_{\rm max}$ is given by $4$ when $\alpha \rightarrow \infty$, for what the threshold saturates.}, $\mathcal{C}_{l,c}$ transition to the type-II region for sufficiently large values of $\kappa$. This illustrates that $\mathcal{C}_c$ does not saturate to a constant value of $\mathcal{C}_c = 2/3$ with $\mathcal{C}_{l,c} = 4/3$ in the type-I region, and we find this is a generic feature for profiles with sufficiently large $\kappa$ values.
Interestingly, we identify a critical point, $\kappa_c \in [30, 40]$, at which $\mathcal{C}_{l,c}$ transitions into the type-II region. However, we find $\kappa_{c}$ to be a quantity dependent on the profile. Our numerical results were obtained for values of $\kappa$ up to approximately $300$. For larger values, the profiles become too sharp, and additional refinement in our numerical simulations would be required to accurately find $\mathcal{C}_{l,c}$. Nevertheless, we expect that the $\mathcal{C}_{l,c}$ does not saturate for even larger values of $\kappa$, though this hypothesis requires careful testing and is left for future research. A top-hat shape in $\zeta_c(r)$ would correspond to the sharpest possible profile ($\kappa \to \infty$), whereas with the metric Eq.\eqref{eq:mtric_K}, the sharpest profile corresponds to a top-hat in $K_{c}(\tilde{r})$ with $(q \to \infty)$ \cite{Escriva:2019phb}, which leads to the maximum value $\mathcal{C}_c = 2/3$ in the type-I region.
The solid blue lines in the figure represent the analytical estimate \cite{Escriva:2019phb} given in Eq.\eqref{eq:delta_c}. The numerical results for $\mathcal{C}_c$ agree very well with the analytical estimate for $\kappa \lesssim 30$ (see the top panel of Fig.\ref{fig:numresults_C_Cl}), but begin to deviate and enter the type-II region for $\kappa \gtrsim 30$. In this regime, $\mathcal{C}_c$ decreases, rather than asymptotically approaching $2/3$. This defines the regime of validity of Eq.\eqref{eq:delta_c} for type-I fluctuations for the shapes $\zeta$ considered. The corresponding value of $q_c$ for this critical $\kappa_c$ is approximately $q_c(\kappa_c = 30) \approx 131.5$. The formula in Eq.\eqref{eq:delta_c} was compared with the numerical results for different profiles of $K(\tilde{r})$ \cite{Escriva:2019nsa, Escriva:2019phb,Escriva:2020tak} up to $q \approx 30$, and therefore our findings are consistent with previous computations with simulations when the threshold lies in the type-I region (which is the case for shapes $K(\tilde{r})$ with the metric Eq.\eqref{eq:mtric_K}).
When comparing the values of $\mathcal{C}_{l,c}$ for the results in the type-II region with the different profiles, we observe that the deviations are on the order of a few percent. This suggests that for the profiles we have used, at least within the regime where our numerical simulations have been done, the profile dependence of $\mathcal{C}_{l,c}$ for $\mathcal{C}_{l}$ is primary sensitive to its curvature $\kappa$ at the maximum $r_m$ (as the case for type-I fluctuations with $\mathcal{C}_{c}$ using the $q$ parameter).

On the other hand, it is important to note that in this regime, the compaction function exhibits a local minimum in the mass excess, surrounded by two local maxima (see Fig.\ref{fig:shapes_C_Cl}). In the limit of very large $\kappa \rightarrow \infty$, this may correspond to an infinitely sharp spike of negative mass excess surrounded by positive regions, which would require further physical understanding. 

\section{Analytical extension to the type-II region}
From the above results, we can extend the analytical estimation of \cite{Escriva:2019phb}, given in Eq.\eqref{eq:delta_c}, to the type-II region using the new parameterization. We use the numerical results from the polynomial curvature profile Eq.\eqref{eq:profile_zeta_pol} to perform this extension. In particular, we define the crossing point $\kappa_{\rm cross} \approx 21.327$, which marks the intersection between the numerical results of Eq.\eqref{eq:profile_zeta_pol} with Eq.\eqref{eq:delta_c}. Then, for $\kappa > \kappa_{\rm cross}$, we fit the data with a non-linear model of the form $\mathcal{C}_{l,c}(\kappa) = a \, \kappa^b$, ensuring that it passes through this crossing point. We obtain $a \approx 0.51962$ and $b \approx 0.26687$. For the cases tested, we find that $\mathcal{C}_{l,c}$ in the large $\kappa$ regime follows a power-law behaviour. Therefore, our analytical formula is as follows: if $\kappa \leqslant \kappa_{\rm cross}$, the functional form of Eq.\eqref{eq:delta_c} applies; otherwise, if $\kappa > \kappa_{\rm cross}$, we use the analytical extension Eq.\eqref{eq:extension_analytical_estimate} to cross to the type-II region.

\begin{align}
    \label{eq:analytical_estimate_deltac}
\delta_{l,c}(\kappa) & = \frac{4}{3}\left( 1 - \sqrt{1-\frac{3}{2} \delta_c(q(\kappa))} \right)  ,\, \, 0 \leqslant \kappa \leqslant \kappa_{\rm cross} \\ \,
\delta_{l,c}(\kappa) &= a \, \kappa ^b , (\,a \approx 0.51962 , \, b \approx 0.26687) , \, \kappa_{\rm cross} <\kappa <300
    \label{eq:extension_analytical_estimate}
\end{align}
where the transcendental relation between $\kappa$ and $q$ is given by Eq.\eqref{eq:s_q} and we denote $\delta_{l,c}(\kappa)$ as the new analytical estimate that mesures $\mathcal{C}_{l,c}$. We can extend the regime of validity of Eq. \eqref{eq:extension_analytical_estimate} for larger $\kappa>300$ as an indication; however, the functional form would need to be tested with new simulations beyond this range. In Appendix~\ref{ref:appendix_skeleton} we show a scheme to analytically estimate the threshold.



\begin{figure} 
\centering
\includegraphics[width=0.45\textwidth]{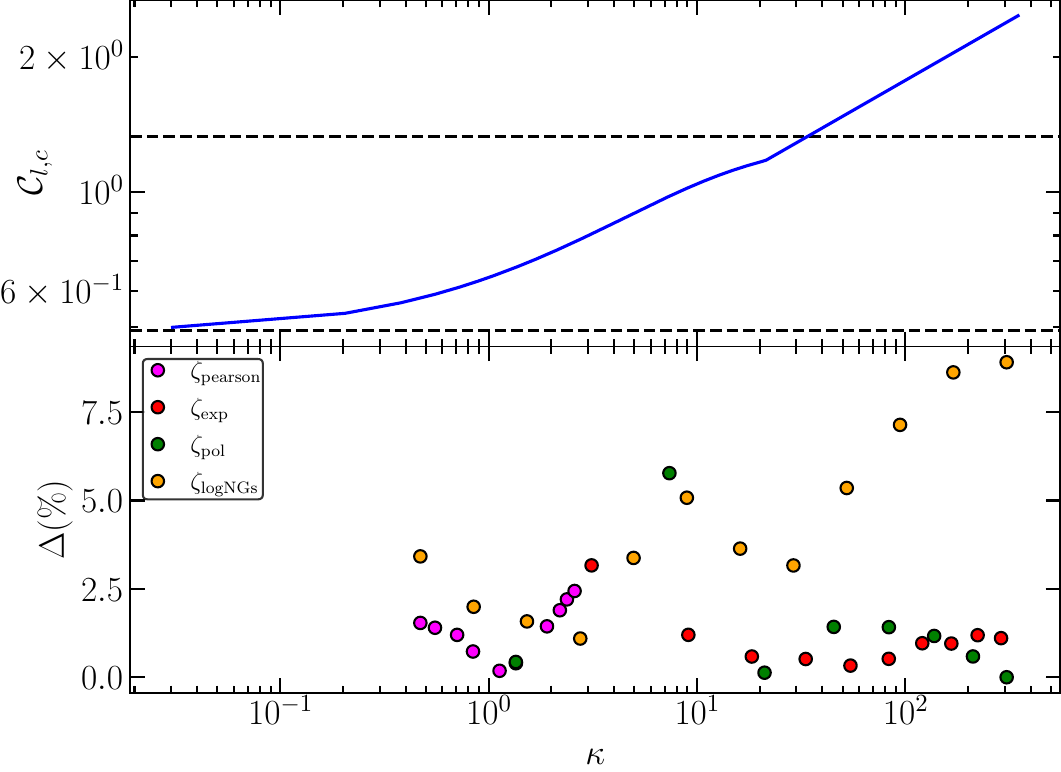}
\caption{Top panel: Analytical estimation Eqs.\eqref{eq:analytical_estimate_deltac} and \eqref{eq:extension_analytical_estimate}. Bottom panel: Relative deviation in percentage between the analytical estimation and the numerical results.}
\label{fig:analytical_plot_formula}
\end{figure}

Finally, in Fig.\ref{fig:analytical_plot_formula}, we show the analytical estimation (top panel) and the relative deviation with respect to the numerical results. The relative deviation is bounded by a few percentage points for the range of $\kappa$ considered, but we find that the shapes of Eq.\eqref{eq:profile_zeta_log} for large negative non-Gaussianity exhibit larger deviations than the other profiles. This may be due to details of the profile around the scale $r_m$, particularly a large negative mass excess region slightly afterward (see Fig.\ref{fig:shapes_C_Cl}). In this regard, it has been shown in \cite{Escriva:2022pnz,Escriva:2023qnq} that the specific details of the shape beyond or before $r_m$ can, to some extent, affect the formation threshold; that is, different functional profile templates with significant differences may lead to different threshold estimates for the same $\kappa$ when compared to each other with a few percent difference.

\section{Conclusions and discussion}

In this work, we have presented results from relativistic numerical simulations with different curvature profiles $\zeta$, exploring the profile dependence of the PBH formation threshold, focusing on type-II fluctuations. Our results show that, for the configurations tested, the threshold for PBH formation in terms of $\mathcal{C}_c$ does not necessarily asymptotically saturate at the boundary between the type-I and type-II regions for sufficiently sharp profiles, contrary to expectations. Instead, we observe a transition of the threshold into the type-II regime. This behaviour appears to be a generic feature, within the range of profiles explored, associated with large curvature in the shape of the linear compaction function $\mathcal{C}_l(r)$ around $r_m$. We quantify this curvature using the dimensionless parameter $\kappa = -r_m^2 \mathcal{C}_l''(r_m)$, and find that $\mathcal{C}_{l,c}(r_m) > 4/3$ (type-II region) when $\kappa > \kappa_c$, with $\kappa_c \in [30, 40]$ for the profiles considered.

Using the numerical results from \cite{MIO} and combining them with the analytical estimation from \cite{Escriva:2019phb}, we have extended the formula Eq.\eqref{eq:delta_c} to estimate $\mathcal{C}_{l,c}$ in the type-II region with $\delta_{l,c}(\kappa)$ Eqs.\eqref{eq:analytical_estimate_deltac} and \eqref{eq:extension_analytical_estimate}. This extension is accurate to within a few percent for the cases tested when compared with the results from the simulations. With our analytical extension into this region, based on the curvature profiles used, more accurate and reliable estimates can be obtained for shapes exhibiting a functional behaviour similar to those considered in our work.

Our findings may have important implications for the statistical estimation of the abundance of PBHs, particularly for models where the threshold of formation lies in the type-II region. In such models, estimations of the PBH production will typically be overestimated due to the expected saturation of $\mathcal{C}_{l,c}$ at the type-I/II boundary. In particular, the critical mean profiles from peak theory \cite{1986ApJ...304...15B}, associated with a monochromatic and a flat spectrum with Gaussian statistics (see \cite{Yoo:2018kvb,Yoo:2020dkz}), correspond to $q \simeq 6$ and $q \simeq 3$, respectively \cite{Musco:2020jjb}, which corresponds to the regime of small $\kappa$ within the type-I region. However, the inclusion of large negative non-Gaussianities (see, for instance, \cite{Pi:2024ert} for such models, and \cite{Pi:2024lsu} for discussions on the importance of non-Gaussianities in the context of PBHs) can modify the shapes such that the critical threshold lies in the type-II region, corresponding to the regime of large $\kappa$. For instance, models with large negative non-Gaussianities have been studied in the context of PBH compatibility with the PTA signal \cite{Franciolini:2023pbf}, showing that the tension present under the assumption of Gaussian fluctuations can be alleviated. Accurate estimates of the collapse threshold are required to compare these models with observational prospects from scalar-induced gravitational waves \cite{Domenech:2021ztg} associated with PBH production, making it important to account for the type-II region. On the other hand, the non-linear statistics developed in \cite{Germani:2019zez,Germani:2023ojx,Fumagalli:2024kxe} provide a valuable framework by integrating over all possible realizations of $\kappa \in (0,\infty)$ peak shapes in the linear compaction function to account for PBH production, following the widely used assumption that type-II fluctuations always collapse to form black holes, and that the threshold therefore does not enter the type-II region. Our results, however, show that the threshold does not necessarily saturate at the type-I/II boundary for sufficiently large $\kappa$. Since the resulting PBH abundance is exponentially sensitive to the collapse threshold, incorporating this effect may be important for obtaining more accurate estimates and for avoiding potential overestimation of PBH production rates for specific realizations with large $\kappa$. 

These findings underscore the relevance of carefully accounting for the type-II region and evaluating its potential impact. Our results provide an important and necessary step toward for completing the critical parameter landscape of PBH formation for type-II fluctuations.


Future directions of our research could involve applying this methodology to scenarios with different equations of state, $w$, thus extending the analytical estimations \cite{Escriva:2020tak} for the type-II region. Additionally, it would be interesting to explore the threshold behaviour for less conventional profiles $\zeta$ and different functional forms than those considered in this work to better understand the threshold behavior for more general shapes, particularly profiles exhibiting specific features such as overlapping compaction function shapes with multiple peaks \cite{Escriva:2023qnq} (see also \cite{Raatikainen:2023bzk}). For such cases, the analysis is expected to be more challenging, and the threshold for collapse may be lower due to the presence of several peaks and surrounding mass excess around $r_m$, as shown in \cite{Escriva:2023qnq}. Moreover, investigating the behaviour in the limit $\kappa \rightarrow \infty$ would be valuable.
\begin{acknowledgments}
{\em  Acknowledgments} 
I thank the support from the YLC program at the Institute for Advanced Research, Nagoya University.
\end{acknowledgments}
\appendix
\section{Convergence of the numerical simulations}
\label{ref:section_convergence}
\begin{figure}[htbp!]
\centering
\includegraphics[width=0.4\textwidth]{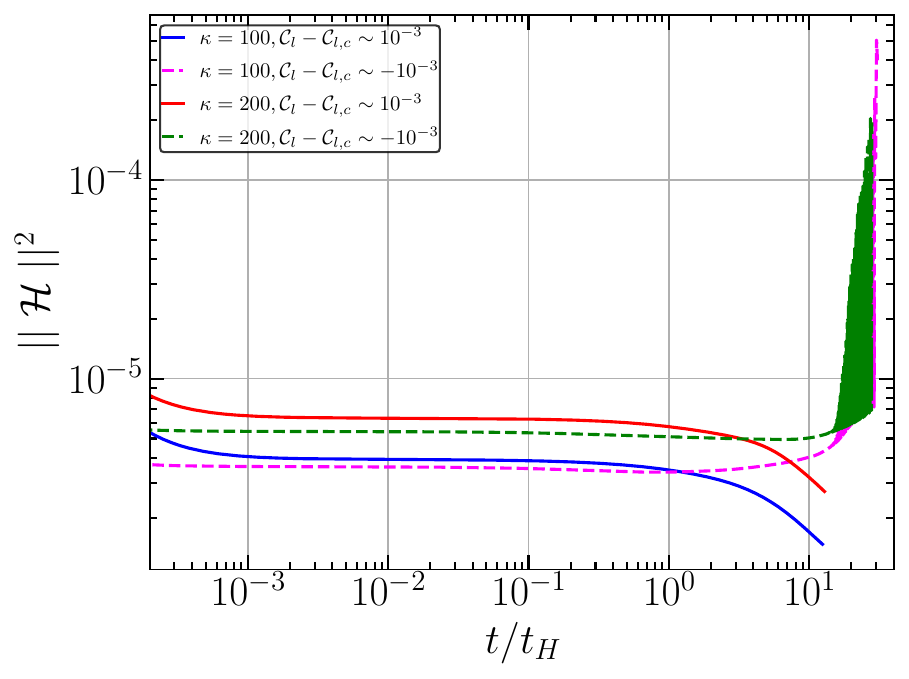}
\includegraphics[width=0.4\textwidth]{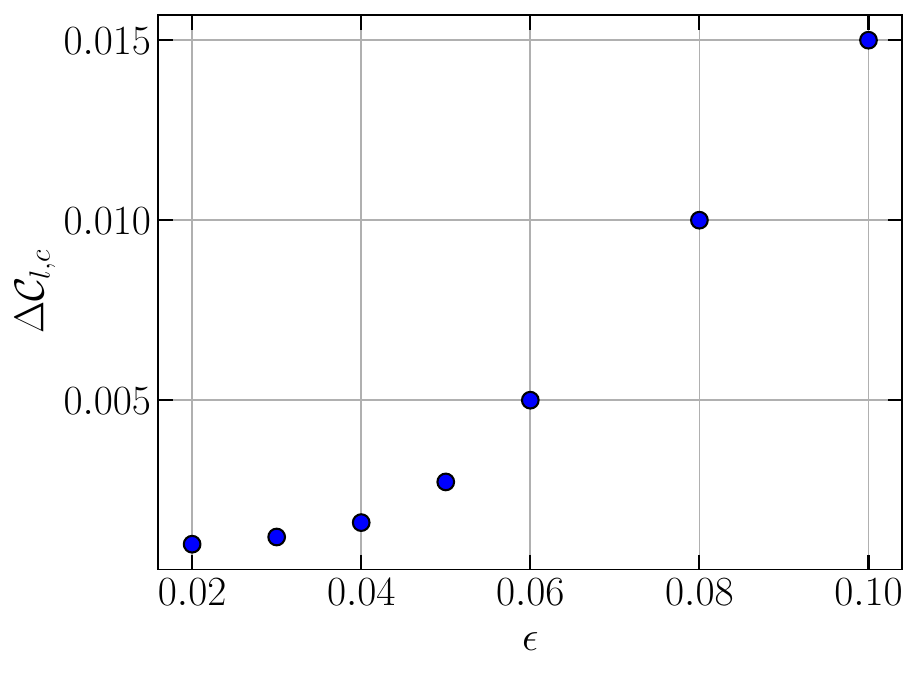}
\caption{Top panel: Numerical evolution of the Hamiltonian constraint for different cases using the exponential profiles Eq.\eqref{eq:profile_zeta_exp} with collapse (solid-lines) and sub-collapse cases (dashed-lines). The \(t_H \equiv \left(\epsilon\, e^{\zeta(r_m)}\right)^{-2}\) corresponds to the time of horizon crossing when the fluctuation reenters the cosmological horizon in a homogeneous background and \(H_0\) is the initial Hubble parameter (see \cite{MIO}). The simulations were initialized with parameters \(t_0 = 1, a_0 = 1\). Bottom panel: Values of \(\Delta \mathcal{C}_{l,c}\) as a function of \(\epsilon\).}
\label{fig:constraint_convergence}
\end{figure}

We demonstrate the convergence of the simulations by analyzing the evolution of the Hamiltonian constraint equations using the exponential profile in Eq.~\eqref{eq:profile_zeta_exp} (similar behavior is observed in other cases), the results are shown in the top panel of Fig.\ref{fig:constraint_convergence}. To test the convergence of truncating the quasi‑homogeneous solution at first order in $\epsilon^2$ for large $\kappa$ values, we perform simulations with various $\epsilon$ and evaluate $\mathcal{C}_{l,c}(\epsilon)$. These values are compared to the reference value $\mathcal{C}_{l,c}(\epsilon=0.02)$ i.e, $\Delta \mathcal{C}_{l,c}(\epsilon) \equiv \mathcal{C}_{l,c}(\epsilon)-\mathcal{C}_{l,c}(\epsilon=0.02)$, which is the used for the results of this work (absolute resolution of $\mathcal{C}_{l,c}$ with $\mathcal{O}(10^{-3})$) at $\epsilon=0.02$. The results are shown in the bottom panel of Fig.\ref{fig:constraint_convergence}.

\section{Skeleton for the threshold analytical determination}
\label{ref:appendix_skeleton}
Here, we provide a general skeleton for obtaining the analytical threshold for PBH formation using Eqs.\eqref{eq:analytical_estimate_deltac} and \eqref{eq:extension_analytical_estimate}. Consider a curvature profile $\zeta(r;\vec{\gamma})$ (with a similar functional behaviour to the profiles of $\mathcal{C}_{l}$ considered), where $\vec{\gamma}$ represents an $M$-dimensional vector of $M$ parameters. To estimate $\mathcal{C}_{l,c}$ using the analytical formula $\delta_{l,c}(\kappa)$, we can follow, for instance, the following steps:

\begin{itemize}
    \item Starting with $\zeta(r; \vec{\gamma})$, obtain the scale $r_m$ which satisfies the equation $\zeta'(r_m; \vec{\gamma}) + r_m \,\zeta''(r_m; \vec{\gamma}) = 0$, for which we have the constraint $r_m \equiv r_m( \vec{\gamma})$;
    
    \item Build the corresponding linear component of the compaction function $\mathcal{C}_l(r) = -4r \zeta'(r)/3$ and obtain the corresponding peak $\mathcal{C}_{l,m} \equiv \mathcal{C}_l(r_m(\vec{\gamma});  \vec{\gamma})$;
    
    \item Fix $M-1$ parameters and relate $\mathcal{C}_{l,m}$ to the remaining one. Iterate over a range of values in $\mathcal{C}_{l,m}$;
    
    \item For each iteration, compute the corresponding $\kappa$ parameter:
    \[
    \kappa = -r_m^2( \vec{\gamma}) \, \mathcal{C}_l'' (r_m( \vec{\gamma}),\vec{\gamma})
    \]
    and introduce the value into $\delta_{l,c}(\kappa)$;
    
    \item If the value $\delta_{l,c}(\kappa)$ does not match the value $\mathcal{C}_{l,m}$ within the desired resolution $\Delta$, proceed with the next iteration of $\mathcal{C}_{l,m}$ until it matches within the desired resolution, i.e., $\delta_{l,c}(\kappa) = \mathcal{C}_{l,m} \pm \Delta$. If $\kappa$ is found to be larger than $\kappa >300$, notice that it goes beyond the range where the analytical estimation has been contrasted with the numerical results.
\end{itemize}

A simple example of a numerical procedure can be found in \cite{codigo_albert}.

\section{Curvature profiles parameterized in terms of $\mathcal{C}_{l}(r_m)$ and $\kappa$}
\label{ref:apendix_shapes_parametrization}
Here we give the curvature profiles of Eq.\eqref{eq:profile_zeta_exp}-\eqref{eq:profile_zeta_log} parametrized in terms of $\mathcal{C}_{l,m} \equiv \mathcal{C}_{l}(r_m)$ and $\kappa$, we define $\tilde{\kappa} \equiv \kappa/\mathcal{C}_{l,m}$.

\begin{align}
\zeta_{\rm exp}(r) &= \frac{3 \, e}{4} \, \frac{\mathcal{C}_{l,m}}{\sqrt{\tilde{\kappa}}} \, \exp(-(r/r_m)^{\sqrt{\tilde{\kappa}}}),\\
\zeta_{\rm pol}(r) &= \frac{3}{\sqrt{2}}\frac{\mathcal{C}_{l,m}}{\sqrt{\tilde{\kappa}}} \frac{ 1}{1+(r/r_m)^{\sqrt{2 \tilde{\kappa}}}},\\
\label{eq:appendix_pearson}
\zeta_{\rm pearson}(r) &=  2^{-3-\frac{8}{(\tilde{\kappa}-4)}} \frac{3 \, \mathcal{C}_{l,m}}{\tilde{\kappa}^{\frac{4}{(4-\tilde{\kappa})}}}\left[ 1-\left( \frac{r}{r_m}\right)^{2} \left( \frac{\tilde{\kappa}-4}{\tilde{\kappa}}\right) \right]^{\frac{\tilde{\kappa}}{\tilde{\kappa}-4}},\\
\zeta_{\rm log NGs}(r) &= -\frac{1}{\lambda(\mathcal{C}_{l,m},\kappa)}\log\left( 1-\lambda(\mathcal{C}_{l,m},\kappa) \,\zeta_G\right), \, \\
\nonumber
\zeta_G &= \mu(\mathcal{C}_{l,m},\kappa) \, \textrm{Sinc}\left(\frac{k_* \, r}{r_m(\mathcal{C}_{l,m},\kappa)}\right).
\end{align}

\bibliographystyle{apsrev4-1}
\bibliography{bibfile}
\end{document}